\documentclass[letterpaper, 10 pt, conference]{ieeeconf}  

\IEEEoverridecommandlockouts                              

\usepackage{graphicx}
\usepackage{rotating}
\usepackage{pdflscape}
\usepackage{array}

\usepackage[table,xcdraw]{xcolor}

\usepackage{dblfloatfix}
\usepackage{fancyhdr}

\fancypagestyle{firstpage}{%
    \fancyhf{}
    \fancyfoot[l]{\scriptsize This work was accepted by the International Conference on Systems, Man, and Cybernetics, Vienna, Austria, 2025. \textcopyright 2025 IEEE. Personal use of this material is permitted. Permission from IEEE must be obtained for all other uses, in any current or future media, including reprinting/republishing this material for advertising or promotional purposes, creating new collective works, for resale or redistribution to servers or lists, or reuse of any copyrighted component of this work in other works.}
    
}

\makeatletter
\newcommand*{\rom}[1]{\expandafter\@slowromancap\romannumeral #1@}
\makeatother

\title{\LARGE \bf
Holistic Specification of the Human Digital Twin: Stakeholders, Users, Functionalities, and Applications
}

\author{Nils Mandischer$^{1,\dagger}$, Alexander Atanasyan$^{2,\dagger}$, Ulrich Dahmen$^{2,\dagger}$, Michael Schluse$^{2}$, \\J\"{u}rgen Rossmann$^{2}$, Lars Mikelsons$^{1}$
\thanks{$^{\dagger}$The three authors contributed equally to this work.}%
\thanks{$^{1}$University of Augsburg, Chair of Mechatronics, 86159 Augsburg, Germany
        {\tt\small nils.mandischer@uni-a.de}}%
\thanks{$^{2}$RWTH Aachen University, Institute of Man-Machine Interaction, 52072 Aachen, Germany
        {\tt\small atanasyan@mmi.rwth-aachen.de}}%
}

\begin{document}
\bstctlcite{IEEEexample:BSTcontrol}

\maketitle
\thispagestyle{empty}
\pagestyle{empty}

\thispagestyle{firstpage}
\begin{abstract}
The digital twin of humans is a relatively new concept. While many diverse definitions, architectures, and applications exist, a clear picture is missing on what, in fact, makes a human digital twin. Within this context, researchers and industrial use-case owners alike are unaware about the market potential of the -- at the moment -- rather theoretical construct. In this work, we draw a holistic vision of the human digital twin, and derive the specification of this holistic human digital twin in form of requirements, stakeholders, and users. For each group of users, we define exemplary applications that fall into the six levels of functionality: store, analyze, personalize, predict, control, and optimize. The functionality levels facilitate an abstraction of abilities of the human digital twin. From the manifold applications, we discuss three in detail to showcase the feasibility of the abstraction levels and the analysis of stakeholders and users. Based on the deep discussion, we derive a comprehensive list of requirements on the holistic human digital twin. These considerations shall be used as a guideline for research and industries for the implementation of human digital twins, particularly in context of reusability in multiple target applications.
\end{abstract}

\section{Introduction}
The concept of the human digital twin (HDT) refers to the digital representation of a person in form of a virtual model. It is based on the idea of capturing, modeling and analyzing data relating to a person's physical, biological, psychological or behavioral aspects, often with real-time constraints. The HDT can serve various purposes, including data integration, simulation and prediction, personalization, and real-time monitoring. The HDT aims to revolutionize the understanding of people and their interactions with their environment. It should not only enable precise decisions and optimized solutions, but also improve the interaction between people and technology in various areas of life. However, the use of an HDT raises questions about data protection, security, and the ethical use of sensitive data. The spectrum of conceivable applications is enormous and ranges from healthcare through the workplace to the education and social sectors. The potential of opportunities can currently only be guessed. However, in terms of practical implementation, there is still a major challenge in translating the huge vision of a holistic HDT into concrete realizations without falling directly into isolated partial solutions. Therefore, a concept is needed for technical implementation while maintaining transferability to other application domains.

In this paper, we will address questions such as:
\begin{itemize}
    \item What could the ideal HDT look like?
    \item What areas of application are relevant?
    \item What boundary conditions need to be considered (e.g., data security, privacy, ethics/acceptance, available standards, available technologies, etc.)?
\end{itemize}
Section~\ref{sec:relatedwork} presents an overview of preliminary work, eventually coming to the conclusion, that the HDT is still in its early stages of development. In Section~\ref{sec:stakeholderapplications}, we present our approach for the systematic identification of potential stakeholders, users, and applications. From this, we derive six levels of functionality required for various applications of the HDT. With the resulting modular framework, we derive a large number of concrete applications. Three of these application proposals are detailed in Section~\ref{sec:applicationsdetail}. Lastly, Section~\ref{sec:specification} lists the comprehensive list of requirements derived from the holistic considerations within this work.

\section{Related Work}
\label{sec:relatedwork}
The concept of the digital twin (DT) has evolved significantly in recent years, becoming a central element in industries such as manufacturing, healthcare, and beyond. However, despite its growing maturity, the field still lacks a universally accepted definition and a common understanding of its core principles~\cite{Naudet2023}. Nevertheless, efforts toward standardization are ongoing, and the development of DTs that can act as ''experimentable'' entities~\cite{Schluse2018} -- digital models that not only mirror their physical counterparts but can also be independently simulated and analyzed -- illustrates the promise of DTs in optimizing systems and processes across a wide range of applications~\cite{Naudet2023}. Building on the foundation laid by DTs, HDTs are already emerging as a transformative concept, particularly in the context of Industry~5.0. HDTs are described as a critical tool for fostering more sustainable, inclusive, and individualized work environments, aligning with Industry~5.0’s human-centric paradigm~\cite{Wang2024}. The potential of HDTs lies in their ability to integrate not only physical and physiological data but also psychological and cognitive aspects, thus enabling a holistic digital representation of the human being. This is considered to pave the way for significant advancements in workplace safety, healthcare personalization, and human-system interaction~\cite{DavilaGonzalez2024}, enabling the holistic modeling of human-to-anything (H2X) systems~\cite{MAS24}.

Academic research on HDTs remains largely focused on conceptual development. Many studies (e.g., by \mbox{Chen et al.~\cite{Chen2013}}) aim to establish frameworks and models that define the components and functionalities an HDT should incorporate. These typically include data acquisition from wearables and IoT devices, coupled with AI-driven models that simulate not only physical attributes but also cognitive and emotional states~\cite{DavilaGonzalez2024}. However, much of the research has not yet progressed beyond the conceptual or prototype phase, with limited real-world implementations~\cite{Loecklin2021}. Where HDT architectures have been developed, they tend to be application-dependent, often tailored for specific industries and applications –- such as narrow treatment response~\cite{Lal2020} and health data visualization~\cite{LauerSchmaltz2023} in healthcare, or intralogistics~\cite{Loecklin2021} and human-robot collaboration~\cite{Montini2021} in manufacturing –- with an emphasis on communication infrastructures and data integration mechanisms~\cite{Okegbile2022}. For instance, studies within the manufacturing domain explore how HDTs can enhance worker safety by monitoring stress levels and other physiological indicators, while others in the healthcare sector propose HDT frameworks aimed at personalized healthcare through real-time data synchronization and AI-based diagnostics~\cite{Okegbile2022,Wang2024}. While in the architectural domain, DTs of buildings are employed as human-building interfaces to balance comfort and energy-saving behavior, DTs of the humans themselves currently appear unconsidered as a possible methodology for optimization and interaction~\cite{Lee2023}.

The literature on legal, ethical, and privacy-related challenges posed by HDTs is underdeveloped, which shows the need for practical realizations of HDTs to not fall behind in the AI race. HDTs involve the collection and analysis of sensitive personal data, including emotional and cognitive metrics, raising concerns around data security, privacy, and worker surveillance~\cite{DavilaGonzalez2024}. While regulations like the General Data Protection Regulation (GDPR) offer a framework for protecting personal data, the application of these frameworks to the vast amount of data processed by HDTs is still underexplored~\cite{Okegbile2022}. Moreover, the ethical implications of continuous human monitoring -— particularly in workplace settings \mbox{-—} warrant further investigation, as balancing worker well-being and privacy is a delicate undertaking~\cite{Wang2024}. Despite these concerns, discussions of regulatory compliance and ethical governance remain limited in the literature. In terms of scalability and integration, another key challenge is the interoperability of HDTs with existing systems (particularly, Brown Field machinery). In industries like healthcare and manufacturing, legacy systems remain widely used, and integrating HDTs into these environments without disrupting established workflows or introducing significant costs is a major hurdle~\cite{Mordaschew2024}. The lack of standardization for data formats and communication protocols further complicates efforts to ensure, that HDTs can seamlessly interact with other systems, including those in manufacturing or health ecosystems~\cite{Naudet2023}. Although some studies have demonstrated prototype architectures for HDTs within specific industrial scenarios, their ability to generalize across broader domains remains limited~\cite{Montini2021}. Despite these challenges, the long-term vision for HDTs extends beyond their current capabilities and limitations. To fulfill their potential, HDTs need to become adaptive and continuously evolving systems that mirror not only the human's current state but also long-term changes in behavior, skills, and health~\cite{Naudet2023}. The technical complexities of maintaining and updating HDTs, particularly in terms of ensuring data accuracy and preventing model obsolescence, are areas that remain underexplored in existing research~\cite{Loecklin2021,Chen2013}. Looking forward, there is a clear need for further exploration into mental and emotional modeling within HDTs~\cite{DavilaGonzalez2024}. Most current implementations focus primarily on physiological data, such as heart rate or fatigue levels, but the integration of psychological factors, including mental well-being, is essential for a more complete representation of the human experience~\cite{Okegbile2022}. Moreover, moving beyond domain-specific solutions towards generalizable holistic HDT frameworks that can be applied across diverse industries, from healthcare to smart buildings, is considered crucial for realizing the full potential of this \mbox{technology~\cite{Wang2024,Montini2021}}.

Summarizing, the development of HDTs is still in its early stages, with the initial progress pointing to significant future possibilities, particularly in the context of Industry~5.0 and healthcare. Overcoming the technical challenges of scalability, interoperability, and long-term maintenance will be essential for moving beyond conceptual models and prototypes to fully functional and widely adopted systems.

\begin{figure*}[t!]
    \centering
    \includegraphics[width=.9\linewidth]{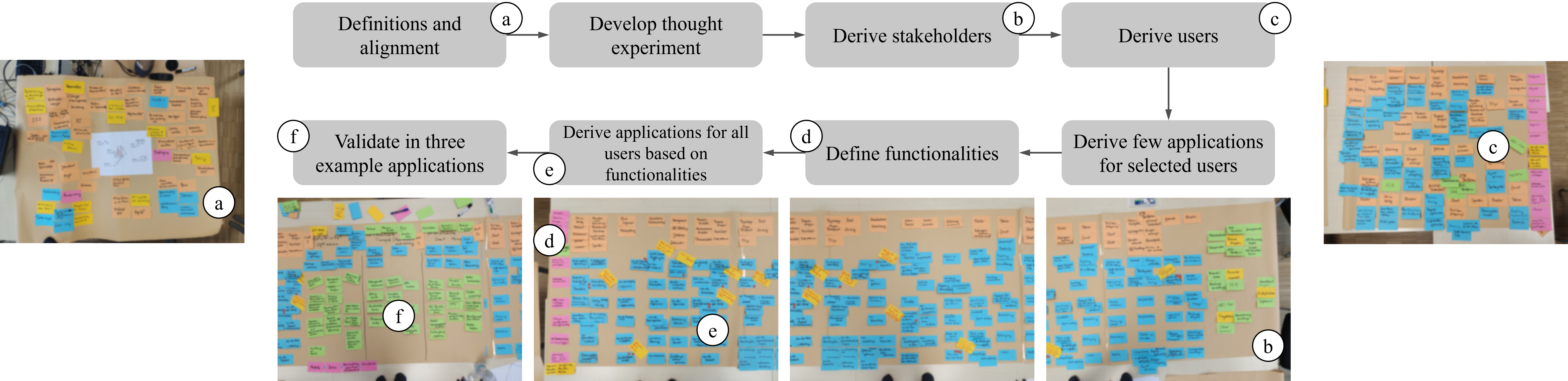}
    \caption{Sketch of the brainstorming approach and examples of the workshop documentation. The workshop language was German.}
    \label{fig:brianstorming}
\end{figure*}

\section{Stakeholders, Users, Functionalities, and Applications}
\label{sec:stakeholderapplications}
To better understand how the HDT needs to evolve in order to meet stakeholder demands, we comprehensively analyze potential stakeholders, users, applications, and required functionalities. From this, we can converge on a shared research vision for the HDT, allowing future efforts for generalization, reuse, and modularity, towards a holistic HDT. We apply a structured brainstorming approach, sketched in Figure~\ref{fig:brianstorming}. The brainstorming group consisted of four technical experts from the domains of modeling and systems engineering, simulation, digital twins, and human-machine interaction. First, a common understanding is established using the Sailboat Retrospective~\cite{Matthies2019} in context of a thought experiment (Section~\ref{ssec:stapp_experiment}) of how a holistic HDT ecosystem may be established in a future society. Second, stakeholders (Section~\ref{ssec:stapp_stakeholders}) and users (Section~\ref{ssec:stapp_users}) are identified. Third, applications are derived for each user and grouped by how they utilize the HDT. From these, we characterize the uses of the HDT to derive levels of functionality (Section~\ref{ssec:stapp_functionality}) needed for its implementation. Fourth, the users are grouped into similar categories. For each of these categories, missing applications are added (Section~\ref{ssec:stapp_applications}), which ultimately results in Table~\ref{tab:hdt_applications_overview} in the Appendix. This step is also used for validation of the levels of functionality. We are able to identify sufficiently diverse applications for each level of functionality and user group, validating the model. Lastly, three specific applications are detailed out, validating the overall framework, which are presented in Section~\ref{sec:applicationsdetail}.

\subsection{Thought Experiment and Conjectures}
\label{ssec:stapp_experiment}
We base the discussion on a thought experiment. Assume that future applications can make use of an idealized HDT in a widespread manner, e.g., a car can load a personalized HDT and customize interior features to the user's needs, or a house could morph in order to be more accessible to the user. As having a user profile exposed in the cloud implies significant security risks, we theorize that such applications would need a more protected way to transfer the configuration of a personalized HDT: The HDT (pre-trained or base configuration) is implemented in the cyber-physical device that shall make use of HDT features. The user has some sort of storage device (e.g., a key card) that carries the personalized information in form of a parametrization of the HDT in an encrypted and protected form. By inserting the storage device into the cyber-physical device, the user acknowledges the usage of the personalized HDT. Such concepts are already in use for critical or personal data, e.g., in form of the electronic patient report (ePA) systems~\cite{AFL11}. The ePA is an example of a very simple HDT that holds health data on an individual. These data are transferred by connecting the health ID card with a terminal, which exposes the data to a physician. In the following considerations, we assume a holistic HDT that functions as an abstract framework. Diverse sources provide models that comply with the HDT framework. Specific applications compose their specific instances of the HDT from a manifold of models and the user data, potentially on edge or in the cloud. The user data serve as the parametrization of the models. Hence, we will only indicate which features an application HDT instance will need to have rather than describing the exact models. By this, we can deduce requirements for a holistic application of the HDT, that should help researchers and practitioners to design their HDTs. By allowing the user to explicitly acknowledge use of their data, we mitigate concerns about safety and security. Note that such a concept will still be subject to strict, particularly, security measures in application (see Table~\ref{tab:requirements_generic} in Section~\ref{sec:specification}).

\subsection{Stakeholders}
\label{ssec:stapp_stakeholders}
A large number of different and diverse actors play a role in such a large-scale ecosystem. As part of our thought experiment, we identify various groups that are relevant to the implementation and operation of an HDT ecosystem. Providers of an HDT infrastructure are key. These include operators of HDT platforms in which the respective HDTs are stored securely, can be accessed at any time, and can be used in a controlled manner. The developers of specific HDT applications and the necessary (embedded) HDT devices (access keys, readers, sensors, displays, etc.) form another group. In addition to these providers of basic functionalities, service providers are conceivable in the context of the HDT ecosystem. These include, e.g., data trustees who take care of the trustworthy transfer of sensitive HDT data, HDT consulting, or certification bodies. Regulatory institutions, such as states, municipalities, and standardization bodies form yet another group. All of these stakeholders and their perspectives must be taken into account when designing and developing a basic framework for the HDT, as all have specific requirements for the HDT. From the stakeholder interests, we later derive requirements that are to be understood as holistic requirements on the HDT. These will need to be specified for individual users that are discussed below.

\subsection{Users}
\label{ssec:stapp_users}
The stakeholders (Section~\ref{ssec:stapp_stakeholders}), who participate in the HDT ecosystem in diverse ways, should be distinguished from the actual users of HDTs. Users are people or groups of people who use their HDT or whose HDT is being used for specific actions, but also people or institutions who use their customers' HDTs. In principle, the group of users is virtually infinite, as every single person can be considered a user, with very specific needs and ideas of what and how their ideal HDT shall be. In our thought experiment, we have therefore selected comprehensive groups of users and application contexts in order to classify potential users. Essentially, this is about how and for what purpose HDTs are used in the respective context. This results in groups such as everyday people, factory workers, teachers, politicians, etc. (see Table~\ref{tab:hdt_applications_overview} in the Appendix). This list of users, although already very detailed, can be extended at will. Our identified users shall give an understanding of the complexity of the HDT ecosystem while sketching potential uses of the HDT to guide its development.

\subsection{Six Levels of Functionality}
\label{ssec:stapp_functionality}
\begin{figure}[tb]
    \centering
    \includegraphics[width=0.43\textwidth]{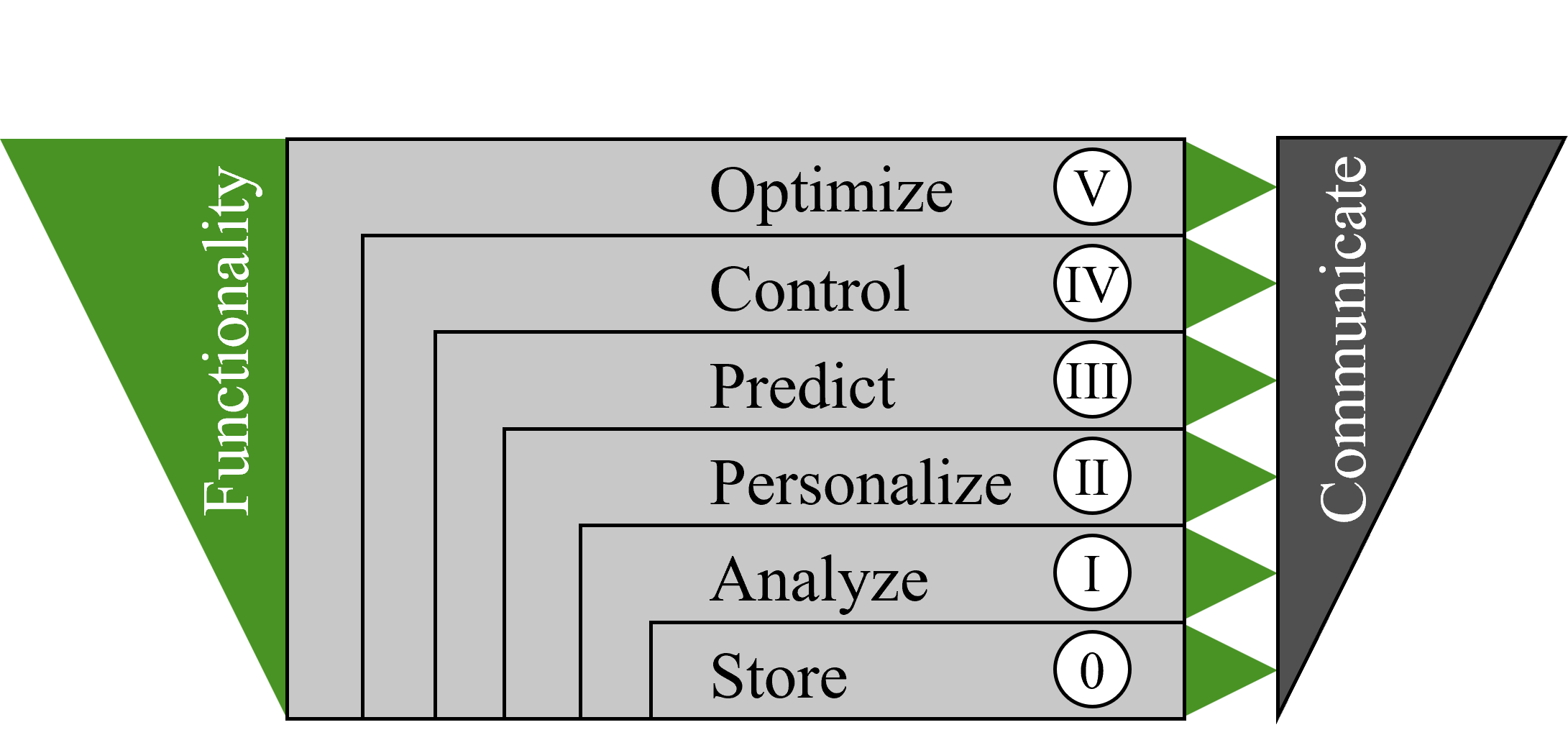}
    \caption{Abstraction of usage/functionality categories. The ``communicate'' functionality is orthogonal to all functionality layers in the HDT.}
    \label{fig:categories}
\end{figure}

\begin{figure}[tb]
    \centering
    \includegraphics[width=0.47\textwidth]{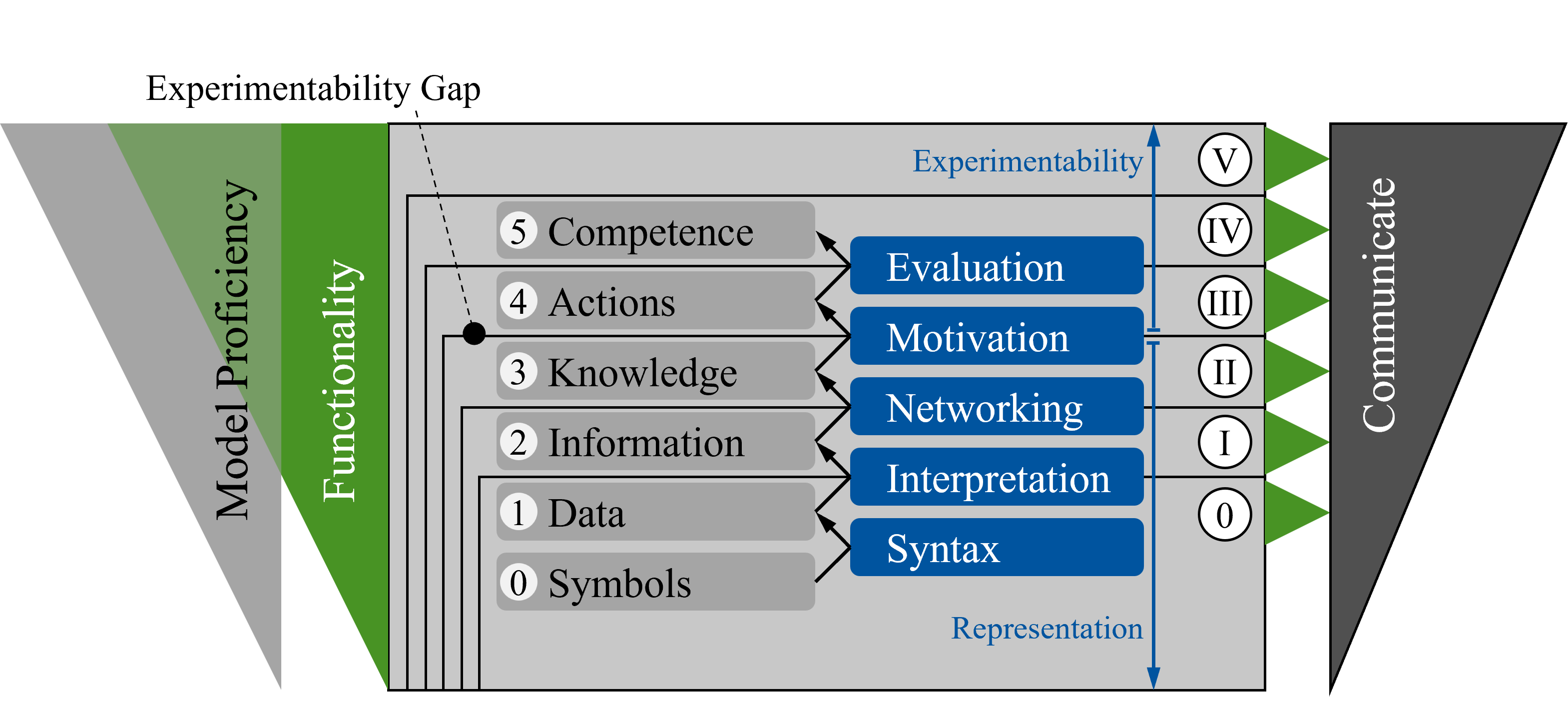}
    \caption{Fractality of the HDT. \emph{Perspectives-Observer-Transparency} model~\cite{MAS24} combined with levels of functionality.}
    \label{fig:fractal}
\end{figure}

For each user, some applications are derived and, consequently, grouped by similar uses. The usage categories were ordered by the amount of functionality required from the HDT. For this, we use a creativity technique: All keywords (including the ones of prior subsections), shall complete the sentence ``The \emph{USER} uses the HDT to \emph{FUNCTIONALITY} \emph{APPLICATION}.'', e.g., ``The \emph{product designer} uses the HDT (of the customer) to \emph{predict user behavior}.'' or ``The \emph{teacher} uses the HDT (of the student) to \emph{personalize the learning pace and plan}.''. Based on this, we derive a comprehensive framework of functionalities. The six resulting levels of functionality of the HDT are:
\begin{enumerate}
    \item[0] \textbf{Store}: HDT used to store data, similar to a data lake. 
    \item[\rom{1}] \textbf{Analyze}: HDT used to evaluate, analyze, or link data.
    \item[\rom{2}] \textbf{Personalize}: HDT used to personalize a cyber-physical entity. Personalization is only performed preemptively, the final configuration is static.
    \item[\rom{3}] \textbf{Predict}: HDT used to predict future data points. A future state is assessed that allows, e.g., validation of a cyber-physical entity (or an idea of it).
    \item[\rom{4}] \textbf{Control}: HDT used to control (feed-forward or backward), modify, or purposefully influence a cyber-physical entity. In contrast to personalization, the configuration may be dynamic during runtime.
    \item[\rom{5}] \textbf{Optimize}: HDT used to optimize a cyber-physical entity with respect to, e.g., costs, time, energy usage, economical gain, or strain.
\end{enumerate}
The levels of functionality are ordered such that lower level functionalities are included in higher level functionality. For example, to personalize a user interface, the HDT first has to assess the state of the human. We also considered the communication of data as a functionality level. However, communication is rather orthogonal to the chosen levels of functionality, as communication takes place on every level. Agglomeration and understandable presentation of data is an important aspect of the HDT to allow communication towards human users, but also to other entities in the HDT ecosystem. The relation of the six levels of functionality and ``communicate'' is depicted in Figure~\ref{fig:categories}. In prior work, we introduced proficiency levels of HDT models within their ability to model, represent, and generate knowledge on the human~\cite{MAS24}. These are analogous to the knowledge development theory by North~\cite{North16}. We find analogy also in the proposed abstraction of the HDT, in which the functionalities can be interpreted as steps towards more proficiency of the HDT. While in~\cite{MAS24}, we describe the way towards competence in understanding the human, in this work, we describe the functionalities (correlating to complexity and number of models) that are needed to implement an application. Both are abstractions of human models, but emphasize different key aspects at different abstraction levels. Here, we define what is required to implement an application and the \emph{Perspectives-Observer-Transparency} model in~\cite{MAS24} gives an ontology on how to combine models in order to achieve this functionality. The combination of both models is depicted in Figure~\ref{fig:fractal}. There seems to be a fractal within the HDT and its models. This could lead to same methods being applicable to diverse scales of the HDT: for specification of functionalities and applications, and for selection and networking of models.

\subsection{Applications}
\label{ssec:stapp_applications}
For all users (Section~\ref{ssec:stapp_users}) and levels of functionality (Section~\ref{ssec:stapp_functionality}), we brainstorm potential applications, in which the users may utilize the HDT. The full list of applications is listed in Table~\ref{tab:hdt_applications_overview} in the Appendix. We omit the level 0 functionality ``store'', as there are virtually infinite applications of data that could be stored.

Prominent examples that already use rudimentary HDTs may be found in the applications. The electronic patient file (ePA) is a digital representation of the human as a patient, storing data on past illnesses, therapies, and medication. The ePA in itself is a typical level 0 functionality. However, depending on with which other HDT models the ePA is composed, more functionality may be gained, e.g., analysis (level I), or preventive functionality (level III). Hence, we consider the ePA to be a good example of how a holistic HDT may be started.
Another prominent example is the language learning app \emph{Duolingo}~\cite{Munday2017}. \emph{Duolingo} manages data on the user, interprets their learning status and behavior, and selects the difficulty of tasks, accordingly. Therefore, \emph{Duolingo} may be classified a level II functionality, or arguably a level IV functionality, as the system -- very rudimentary -- learns from how users interact with the language tasks. These two examples can be found in multiple of the identified applications, as we observed that applications within each level of functionality appear to have similarities, particularly on the lower levels. Hence, inherent functionality of these examples is shared with many other applications.

\section{Applications in Detail}
\label{sec:applicationsdetail}
To validate the functionalities in Section~\ref{ssec:stapp_functionality} and the applications in Section~\ref{ssec:stapp_applications}, we choose three representative applications and detail them, defining users, uses/main functionalities of the HDT, and required functionality (models, data). The profiles of these applications are listed in Table~\ref{tab:detailedapps}.

\subsection{Individualized Therapy/Administering Medication}
\label{ssec:appdetail_therapy}
In individualized therapy, the HDT is used in the process of planning and administering medication. This requires the HDT to predict the outcomes of the selected medication or treatment via respective models, and to transfer the predictions to the patient data, e.g., to  provide information about possible individual side effects. We sketch the system as a decision support system. Hence, the HDT does not have an immediate feedback loop to learn from the administered medication, classifying it as a level III functionality.  

\subsection{Smart Home}
\label{ssec:appdetail_smarthome}
In a building, the HDT may be used to personalize a smart home and its connected devices. By accessing the data of the user's (here: the tenant or house owner) HDT, the smart home system can be enhanced with more data, models, and opportunities than conventional smart home services. In this application, the HDT is used to personalize a smart home experience to the inhabitant, by not only providing data and models but directly controlling the devices. As the HDT itself relies on its embedding platform to control other entities, it must implement the necessary interfaces, e.g., to a smart home controller. Therefore, the main task of the HDT is to model and simulate the person's interaction with the building/home environment, e.g., through comfort models (temperature, sound, etc.). Through experimentability, multiple scenarios may be simulated and conclusions taken on the exact actions to improve the performance indices given by the usage (Table~\ref{tab:detailedapps}). Due to legal regulations (e.g., GDPR) or personal preferences, the HDT may be prohibited from using cameras and microphones as sensing modalities. Although the scenario title suggests a classification as Level II functionality, the ability to set and control device parameters corresponds to Level IV functionality. This distinction is reflected in Table~\ref{tab:detailedapps} by the increased number of required functions compared to the previous application. Furthermore, the communication complexity increases: Rather than showing the physician information on a medication decision, the HDT must communicate on multiple levels (human-machine interfaces, user interfaces, machine communication protocols, etc.), and with multiple devices and users.
\begin{table}[tb]
    \newcommand{\hyphentemp}{\hyphenpenalty=10\exhyphenpenalty=10} 
    \newcommand{\ghline}{\arrayrulecolor{gray}\hline}   
    \newcommand{\bhline}{\arrayrulecolor{black}\hline}  
    \newcommand{\redfs}{\footnotesize}
    \centering
    \caption{Profiles of the three detailed applications.}
    \label{tab:detailedapps}
    \begin{tabular}{*{3}{>{\hyphentemp}p{.28\linewidth}}}
        \bhline
        \rowcolor[gray]{0.7} \multicolumn{3}{c}{\textbf{Level III: Administering Medication}}\\
            \bhline
            \rowcolor[gray]{0.9} \multicolumn{3}{c}{\textbf{Users}}\\
            \bhline
            {\redfs patient} & {\redfs physician} & {\redfs nurse}\\
            \ghline
            {\redfs health insurer} & \multicolumn{2}{l}{\redfs pharmaceutical company}\\
            \bhline
            \rowcolor[gray]{0.9} \multicolumn{3}{c}{\redfs \textbf{Uses of the HDT}}\\
            \bhline
            {\redfs get/stay healthy} & {\redfs safe costs} & {\redfs reduce interactions}\\
            \bhline
            \rowcolor[gray]{0.9} \multicolumn{3}{c}{\redfs \textbf{Required Functionality}}\\
            \bhline
            {\redfs evaluate condition} & {\redfs medication data base} & {\redfs interface w. ePA}\\
            \ghline
            {\redfs evaluate well-being} & {\redfs interface w. sensors} & {\redfs human behavior}\\
            \ghline
            {\redfs medication interactions} & {\redfs evaluating medication effects} & {\redfs track progress of therapy}\\
            \ghline
            \multicolumn{2}{l}{\redfs in silico medication tests} & ~\\
        \bhline
        \bhline
        \rowcolor[gray]{0.7} \multicolumn{3}{c}{\textbf{Level IV: Smart Home (SH)}}\\
            \bhline
            \rowcolor[gray]{0.9} \multicolumn{3}{c}{\textbf{Users}}\\
            \bhline
            {\redfs landlord} & {\redfs inhabitants} & {\redfs smart home provider}\\
            \bhline
            \rowcolor[gray]{0.9} \multicolumn{3}{c}{\textbf{Uses of the HDT}}\\
            \bhline
            {\redfs safe energy/costs} & {\redfs control light} & {\redfs improve comfort}\\
            \ghline
            {\redfs safe time} & {\redfs control SH} & {\redfs increase level of automation}\\
            \bhline
            \rowcolor[gray]{0.9} \multicolumn{3}{c}{\textbf{Required Functionality}}\\
            \bhline
            {\redfs interface with SH} & {\redfs evaluate comfort} & {\redfs control SH devices}\\
            \ghline
            {\redfs interface w. weather data} & {\redfs predict user behavior} & {\redfs interface w. schedules}\\
            \ghline
            {\redfs learn from interaction w. user} & {\redfs model control problem} & {\redfs model body temperature}\\
            \ghline
            {\redfs model light sensation} & {\redfs GDPR conformity} & ~\\
        \bhline
        \bhline
        \rowcolor[gray]{0.7} \multicolumn{3}{c}{\textbf{Level V: Optimize Factory Productivity}}\\
            \bhline
            \rowcolor[gray]{0.9} \multicolumn{3}{c}{\textbf{Users}}\\
            \bhline
            {\redfs worker} & {\redfs management} & {\redfs layout designer}\\
            \ghline
            {\redfs engineer} & {\redfs work preparation} & {\redfs software provider}\\
            \bhline
            \rowcolor[gray]{0.9} \multicolumn{3}{c}{\textbf{Uses of the HDT}}\\
            \bhline
            {\redfs improve ergonomics} & {\redfs improve motivation} & {\redfs improve layout}\\
            \ghline
            {\redfs improve worker performance} & {\redfs improve overall productivity} & {\redfs increase profit/turnover}\\
            \ghline
            {\redfs reduce absenteeism} & {\redfs distribute workers} & ~\\
            \bhline
            \rowcolor[gray]{0.9} \multicolumn{3}{c}{\textbf{Required Functionality}}\\
            \bhline
            {\redfs evaluate indiv. performance} & {\redfs conformity with works council} & {\redfs model optimization problem}\\
            \ghline
            {\redfs GDPR conformity} & {\redfs experimentability} & {\redfs evaluate HMI}\\
            \ghline
            {\redfs model complete work system} & {\redfs predict possible futures} & {\redfs evaluate possible futures}\\
            \ghline
            {\redfs interface with objects, machines, environment} & {\redfs interface with design software} & {\redfs interface with Brown Field}\\
            \ghline
            {\redfs evaluate ergonomics} & \multicolumn{2}{l}{\redfs control ergonomics parameters} \\
            \bhline
    \end{tabular}
\end{table}
\begin{table*}[b!]
    \newcommand{\hyphentemp}{\hyphenpenalty=10\exhyphenpenalty=10} 
    \caption{Generic requirements on the HDT.}
    \label{tab:requirements_generic}
    \centering
    \begin{tabular}[t]{p{.008\linewidth} | >{\hyphentemp}p{.172\linewidth} >{\hyphentemp}p{.74\linewidth}}
        \hline
        \rowcolor[gray]{0.9} \textbf{A} & \multicolumn{2}{l}{\textbf{Safety \& Security}}\\
        \hline
        1 & User safety & The HDT within any configuration shall not harm the user, groups of users, or the society. \\
        2 & Digital sovereignty & The user must be in control of their data at all times. This includes explicit acknowledgement of data use, protection against data theft, and the form of data transfer (blocked, anonymized, pseudonymized). \\
        3 & Transparency & Clarity about which data is used and how. \\
        4 & Persistence & The HDT must be fail-safe, i.e. current and history data must not be lost in an uncontrolled manner. \\
        5 & Traceability &  The HDT shall maintain a history of all data and its changes (versioning). \\
        6 & Verifiability & The authenticity of an HDT and the correctness of the HDT's data must be verifiable (no fake or dummy identity). \\
        7 & Data consistency & Data within the HDT must be consistent and unique (single source of truth), to be secured against data corruption. \\
        \hline
        \rowcolor[gray]{0.9} \textbf{B} & \multicolumn{2}{l}{\textbf{Reliability}}\\
        \hline
        1 & Maintaina\-bility & The HDT as a whole must be maintainable.\\
        2 & Fault tolerance & The HDT must not be brought into an uncontrollable state by any configuration. When the HDT enters an uncontrollable or unintended state, it shall return to a controllable save state (risk/fault mitigation). \\
        3 & Zero obsolescence & The HDT must maintain compatibility with legacy implementations of HDT system elements at all time. The HDT must not cease function.\\ 
        \hline
        \rowcolor[gray]{0.9} \textbf{C} & \multicolumn{2}{l}{\textbf{Legal \& Regulatory}}\\
        \hline
        1 & Compliance w. international law & The HDT must comply with the common ground in international law (part. according to the largest markets in Europe, Northern America, and East Asia), and laws that are considered good practice. \\
        2 & Adaptation to local law \& culture & The HDT must be adaptable to local legal and cultural conditions, also featuring the local understanding of ethics. \\
        \hline
        \rowcolor[gray]{0.9} \textbf{D} & \multicolumn{2}{l}{\textbf{Structure \& Functionality}}\\
        \hline
        1 & Maintain diverse models & Hold and operate models and data of diverse nature, level of detail, and granularity.\\
        2 & Model composition & Allow composition of models w. models, models w. data, and data w. data. \\
        3 & Model selection & Allow selection from all models, including model alternatives (e.g., of different manufacturers), based on desired higher-level function. \\
        4 & Singleton & Guarantee that there exist only one unique HDT for one person (``One Person, One HDT''). \\
        5 & Extendability & Models and data of the HDT must be extendable. \\
        6 & Modularity & The HDT shall be extendable to new use cases and specific applications should be independent of each other. \\
        7 & Adapt to specific human & Allow configuration/paramaterization of the HDT to a specific human. \\
        8 & Plannable death & Allow the HDT to be completely purged (incl. all data), if desired, including to specific dates or events. \\
        \hline
        \rowcolor[gray]{0.9} \textbf{E} & \multicolumn{2}{l}{\textbf{Connectivity \& Interchangeability}}\\
        \hline
        1 & Independence of service providers & Users of the HDT must be free to choose their service providers that host the HDT.\\
        2 & Interopera\-bility w. HDTs & The HDT must be interoperable with other HDTs, includes compliance with DT standards.\\
        3 & Interopera\-bility w. assets \& services & The HDT must be interoperable with assets and services (Green Field). \\
        \hline
        \rowcolor[gray]{0.9} \textbf{F} & \multicolumn{2}{l}{\textbf{Accessibility}}\\
        \hline
        1 & Participation & Everyone must be able to participate in the HDT ecosystem, esp. having the ability to get their personalized HDT. \\
        2 & Human understandable I/O & The input and output interfaces of the HDT must be understandable to a broad audience in a natural way. \\
    \end{tabular}
\end{table*}

\subsection{Human-Centered Productivity Optimization}
\label{ssec:appdetail_productivity}
In factories, the HDT may be used to optimize the productivity of human-operated workstations. Particularly in linear production, humans work alongside automation, and an increasing number of tasks is allocated to automation systems. The leftover manual workplaces, i.e. those uneconomical for automation, usually feature repetitive labor and require either fine motor functions, precise handling in presence of obstacles, or machine operation. These -- often dull -- tasks lead to declining motivation of the workers. We foresee the HDT to be used as a measure to optimize the interaction of the workers and their environment to optimize the environment for worker motivation and well-being, which leads to reduced absenteeism and an overall improved productivity. This endeavor requires manifold and complex functionality, particularly in assessing hidden human states (e.g., well-being), and diverse communication strategies. This system is not only subject to GDPR, but also to company- or factory-specific culture and regulations, e.g., implemented by a works council or union. As we target the HDT to be used for optimization of the working environment, it requires level V functionality. Optimization does not only require to react to control variables, but also to learn from interactions between the HDT and the interaction context, and to adapt its own configuration accordingly, to converge to optimized states w.r.t. defined metrics. Again, the number of required functions increases alongside the level of functionality compared to the other two applications, implying a linear or superlinear relation between the complexity of the main functionality and the number of models, data, sub-systems, and system elements. Another observation is that not only the complexity of functionality and communication rises with the level of functionality, but also the legal and regulatory requirements.

\section{Specification of the HDT}
\label{sec:specification}
From the comprehensive definition of the HDT, its users, stakeholders, and applications in Sections~\ref{sec:stakeholderapplications} and \ref{sec:applicationsdetail}, we derive holistic requirements on the HDT in any application. The requirements (Table~\ref{tab:requirements_generic}) are derived in a three-step sequential brainstorming process: (1) derive requirements from definitions in Sections~\ref{sec:stakeholderapplications} and \ref{sec:applicationsdetail}, (2) group requirements, define categories, and uncover redundancies, (3) discuss each category of requirements and uncover missing requirements.

\textbf{Safety \& Security~(A)} lists all requirements that shall keep the user and their data safe, and which secure the HDT and guarantee its internal consistency. Data security is not explicitly listed as it is a necessary solution to digital sovereignty (A2). Further, user safety (A1) includes a holistic understanding of users, also establishing methods for safety of users, user groups, and society, similar to how Asimov prioritizes safety in his laws of robotics~\cite{Asimov1950}.

\textbf{Reliability~(B)} features requirements that shall guarantee the availability of the HDT. We choose not to list low obsolescence, i.e. that implementations of the HDT shall not become obsolete but may become in the long run, but to list the more restrictive requirement for zero obsolescence, i.e. implementations of the HDT must never become obsolete. In light of the thought experiment (Section~\ref{ssec:stapp_experiment}), the HDT shall become an ever-present entity with the majority of the population participating within its ecosystem (see item F1). Such a large scope requires all kind of HDT implementations to allow participation, hence, they must not become obsolete or have fall-back plans to update to non-obsolete versions in case of planned obsolescence, i.e. HDT implementations may only become obsolete if no instance is in use anymore.

\textbf{Legal \& Regulatory~(C)} lists requirements guaranteeing that the HDT satisfies regulatory boundary conditions. Such boundary conditions may come from local or international legislation or local culture and conventions.

\textbf{Structure \& Functionality~(D)} features comprehensive functional and structural requirements of the software entity HDT. The HDT is a model and data warehouse composed of diverse models of human sub-systems and system elements, also featuring similar models for the same sub-entity, e.g., multiple approaches to model human gait. Model selection~(D3) on the one hand, is the selection of models required to fulfill a main functionality as required by the application, which also features selecting from a manifold of similar models to achieve the best possible performance. Model composition (D2), on the other hand, is the integration of multiple models into higher-level models, guaranteeing cross-functionality. We further foresee there to be only a single instance of an HDT for an individual person (D4). This singleton HDT has significant implications on its longevity, particularly given the mortality of its owner. Plannable death~(D8) allows the owner to create a will on the HDT, controlling if the HDT shall be purged after death or if it shall live on in the cyberspace, but also to give the owner control if the HDT shall be purged prematurely.

\textbf{Connectivity \& Interchangeability (E)} lists requirements guaranteeing the connectivity of the HDT with entities inside~(E2) and outside (E3) the HDT ecosystem. Interchangeability implies independence of service providers hosting the HDT (E1). We foresee the HDT to not be completely computable on the edge or on-chip, hence, hosting services will be required. These requirements imply that there needs to be a shared standard for HDTs and that the HDT complies with existing DT standards, e.g., \mbox{ISO/IEC 30173:2023}~\cite{ISO30173}, to guarantee connectivity to Green Field devices.

\textbf{Accessibility~(F)} features all requirements allowing people, particularly in context of diversity, to participate in the HDT ecosystem. While E features accessibility by cyber-physical systems, F features accessibility by the human. The \emph{Universal Design} paradigm (F3) allows any people to use a product or participate in a target system without further customization~\cite{Goldsmith2000}. The European Accessibility Act~\cite{AccessibilityAct} is currently in the process of being implemented into local law and requires new products to be universally designed. Another important aspect which substantially determines the success of the HDT is the economic viability (F4). In a scope as large as the HDT spans, engineers shall focus on what is required to fulfill the main functionality in context of the application, and not implement or configure additional functionality that potentially renders the HDT uneconomical.

In conclusion, Table~\ref{tab:requirements_generic} draws a holistic vision of the HDT that can only be met with demanding requirements. However, if a holistic HDT shall be implemented, the large scope has to be considered, but individual requirements may become less complex in application as, e.g., only few models, data, or interfaces are required for the target application (cf. Section~\ref{sec:applicationsdetail}). We hope these requirements will help engineers to guide implementation of HDTs and to maintain a perspective on the larger scope of human modeling.

\section{Conclusion}
In this work, we derived manifold details on the HDT and its application. First, a holistic analysis was performed to derive stakeholders, users, and potential applications based on a thought experiment. In our vision, the HDT is a holistic representation of the human that is able to be embedded into arbitrary hardware assets of everyday life. Based on this analysis, we derived six levels of functionality, offering an eased representation of complexity of an HDT implementation required for specific applications. Again, based on the holistic analysis, we derived comprehensive and generic requirements on the HDT, to guide implementation and research on the HDT. With these items, a guideline specification of a holistic HDT is given.

\addtolength{\textheight}{-12cm}   


\section*{APPENDIX}
Table~\ref{tab:hdt_applications_overview}: All applications derived for each functionality and user in Section~\ref{sec:stakeholderapplications}.




\bibliographystyle{IEEEtran}
\bibliography{references}

\begin{thebibliography}{10}
\providecommand{\url}[1]{#1}
\csname url@samestyle\endcsname
\providecommand{\newblock}{\relax}
\providecommand{\bibinfo}[2]{#2}
\providecommand{\BIBentrySTDinterwordspacing}{\spaceskip=0pt\relax}
\providecommand{\BIBentryALTinterwordstretchfactor}{4}
\providecommand{\BIBentryALTinterwordspacing}{\spaceskip=\fontdimen2\font plus
\BIBentryALTinterwordstretchfactor\fontdimen3\font minus \fontdimen4\font\relax}
\providecommand{\BIBforeignlanguage}[2]{{%
\expandafter\ifx\csname l@#1\endcsname\relax
\typeout{** WARNING: IEEEtran.bst: No hyphenation pattern has been}%
\typeout{** loaded for the language `#1'. Using the pattern for}%
\typeout{** the default language instead.}%
\else
\language=\csname l@#1\endcsname
\fi
#2}}
\providecommand{\BIBdecl}{\relax}
\BIBdecl

\bibitem{Naudet2023}
Y.~Naudet, C.~Stah, and M.~Gallais, ``Preliminary systemic model of (human) digital twin,'' in \emph{PErvasive Technologies Related to Assistive Environments}, 2023.

\bibitem{Schluse2018}
M.~Schluse, M.~Priggemeyer, L.~Atorf, and J.~Rossmann, ``Experimentable digital twins—streamlining simulation-based systems engineering for industry 4.0,'' \emph{IEEE Transactions on Industrial Informatics}, vol.~14, 2018.

\bibitem{Wang2024}
B.~Wang \emph{et~al.}, ``Human digital twin in the context of industry 5.0,'' \emph{Robotics and Computer-Integrated Manufacturing}, vol.~85, 2024.

\bibitem{DavilaGonzalez2024}
S.~Davila-Gonzalez and S.~Martin, ``Human digital twin in industry 5.0: A holistic approach to worker safety and well-being through advanced ai and emotional analytics,'' \emph{Sensors}, vol.~24, 2024.

\bibitem{MAS24}
N.~Mandischer, A.~Atanasyan, M.~Schluse, J.~Roßmann, and L.~Mikelsons, ``Perspectives-observer-transparency – a novel paradigm for modelling the human in human-to-anything interaction based on a structured review of the human digital twin,'' in \emph{IEEE International Conference on Systems, Man, and Cybernetics (SMC)}, 2024.

\bibitem{Chen2013}
J.~Chen, C.~Yi, S.~D. Okegbile, J.~Cai, and X.~Shen, ``Networking architecture and key supporting technologies for human digital twin in personalized healthcare: A comprehensive survey,'' 2023.

\bibitem{Loecklin2021}
A.~Löcklin, T.~Jung, N.~Zazdi, T.~Ruppert, and M.~Weyrich, ``Architecture of a human-digital twin as common interface for operator 4.0 applications,'' \emph{Procedia CIRP}, vol. 103, 2021.

\bibitem{Lal2020}
A.~Lal \emph{et~al.}, ``Development and verification of a digital twin patient model to predict specific treatment response during the first 24 hours of sepsis,'' \emph{Crit Care Explor.}, vol.~2, 2020.

\bibitem{LauerSchmaltz2023}
M.~W. Lauer-Schmaltz, I.~Kerim, J.~P. Hansen, {Gábor Máté Gulyás}, and H.~B. Andersen, ``Human digital twin-based interactive dashboards for informal caregivers of stroke patients,'' in \emph{PErvasive Technologies Related to Assistive Environments}, 2023.

\bibitem{Montini2021}
E.~Montini \emph{et~al.}, \emph{Trusted Artificial Intelligence in Manufacturing: A Review of the Emerging Wave of Ethical and Human Centric AI Technologies for Smart Production}.\hskip 1em plus 0.5em minus 0.4em\relax now, 2021.

\bibitem{Okegbile2022}
S.~D. Okegbile, J.~Cai, D.~Niyato, and C.~Yi, ``Human digital twin for personalized healthcare: Vision, architecture and future directions,'' \emph{IEEE Network}, vol.~37, 2022.

\bibitem{Lee2023}
K.~S. Lee, J.-J. Lee, C.~Aucremanne, I.~Shah, and A.~Ghahramani, ``Towards democratization of digital twins: Design principles for transformation into a human-building interface,'' \emph{Building and Environment}, vol. 244, 2023.

\bibitem{Mordaschew2024}
V.~Mordaschew, S.~Duckwitz, and S.~Tackenberg, ``A human digital twin of disabled workers for production planning,'' in \emph{5th International Conference on Industry 4.0 and Smart Manufacturing}, 2024.

\bibitem{Matthies2019}
C.~Matthies, F.~Dobrigkeit, and A.~Ernst, ``Counteracting agile retrospective problems with retrospective activities,'' in \emph{Systems, Software and Service Process Improvement}, 2019.

\bibitem{AFL11}
N.~Archer, U.~Fevrier-Thomas, C.~Lokker, K.~A. McKibbon, and S.~E. Straus, ``Personal health records: a scoping review,'' \emph{Journal of the American Medical Informatics Association}, vol.~18, no.~4, 07 2011.

\bibitem{North16}
K.~North, ``Die wissenstreppe,'' \emph{Wissensorientierte Unternehmensf\"{u}hrung}, 2016.

\bibitem{Munday2017}
P.~Munday, ``Duolingo. gamified learning through translation,'' \emph{Journal of Spanish Language Teaching}, vol.~4, 2017.

\bibitem{Asimov1950}
I.~Asimov, \emph{I, Robot}.\hskip 1em plus 0.5em minus 0.4em\relax Gnome Press, 1950.

\bibitem{ISO30173}
{ISO/IEC 30173:2023}, ``Digital twin - concepts and terminology,'' 2023.

\bibitem{Goldsmith2000}
S.~Goldsmith, \emph{Universal Design}.\hskip 1em plus 0.5em minus 0.4em\relax {Taylor \& Francis Group}, 2000.

\bibitem{AccessibilityAct}
``{Directive (EU) 2019/882 of the European Parliament and of the Council of 17 April 2019 on the accessibility requirements for products and services},'' Official Journal of the European Union, 2019.

\end{thebibliography}

\addtolength{\textheight}{12cm}
\clearpage
\setlength{\tabcolsep}{3pt}
\begin{sidewaystable}[t]
\vspace{25em} 
\caption{Applications of the HDT.}
\label{tab:hdt_applications_overview}
{
    \begin{minipage}{\textwidth}
        \newcommand{\hyphentemp}{\hyphenpenalty=0\exhyphenpenalty=0} 
        \small
        \resizebox{1.00\textwidth}{!}{%
        \begin{tabular}{>{\hyphentemp}>{\raggedright\arraybackslash\hspace{0pt}}m{2cm}|*{16}{>{\hyphentemp}>{\raggedright\arraybackslash\hspace{0pt}}m{2cm}}}
            \textbf{Function} & \textbf{Me in Everyday Life} & \textbf{Work Planner} & \textbf{Shift Supervisor} & \textbf{Management} & \textbf{Product Manufacturer} & \textbf{Psychologist} & \textbf{(Health) Insurance} & \textbf{Teacher, Trainer, Learner} & \textbf{Vehicle} & \textbf{Military} & \textbf{Police, Lawyer} & \textbf{Politician} & \textbf{Trade} & \textbf{Buildings} & \textbf{Researcher} \\
            \hline
            \textbf{Analyze \mbox{(Level~I)}} & nutrition (planning); self-diagnosis; health monitoring & human/team performance & current state; human/team performance & screen impostors; find candidates for jobs & user needs; usage behavior; mass customization & health (remote diagnosis); reimbursements (doctor, manufacturer, patient) & profiling; intake assessment; reimbursements (doctors, manufacturers, patient) & continuous skill tracking; potential; learning progress; emotional state & driver’s condition & monitor soldier’s condition; allergies/medication in emergencies & credibility; forensics; systematic profiling; evidence & population statistics; single source of truth (citizen data) & assortment; energy consumption behavior; demand estimation & usage behavior & generate new insights into unknown phenomena \\
            
            \textbf{Personalize \mbox{(Level~II)}} & adaptation of working environment; assistance; individual support & environment (ergonomics) & work plan; skill-based assignments & operator-independent machinery & end-user customization; batch size one; individualized product; interaction with product & personalized therapy; medication & individualized fees & personal learning pace & car-sharing: personalized settings/ configuration; driving style; safety features (e.g., for children) & personalized equipment & personalized equipment & election advertising; microtargeting & advertising; microtargeting; multi-level marketing & adaptation of buildings (heating, elevators, etc.) & tailored methods \\
            
            \textbf{Predict \mbox{(Level~III)}} & predictive human maintenance (preventive healthcare); safety/ security (actions) & user behavior validation & performance & performance & product/system behavior; target specifications; user behavior prediction & therapy outcome (predictive); disease (predict); predictive human maintenance & predictive human maintenance & learning success & driver’s future state & identify intentions; identify targets & forensic analysis/crime reconstruction; validate investigations & validate legislation & purchasing behavior & predict occupants' behavior; validation & evaluate methods; generate hypotheses \\
            
            \textbf{Control \mbox{(Level~IV)}} & smart home: room/device adjustment; environment control; increase accessibility; nutritional planning & product line & team composition; optimal tasks for each worker & reduce absenteeism & product behavior; product adaptation & influence patient behavior; disease (prevent) & treatment procedures & learning steps; introduce empathy into app; mood regulation (stress management) & self-control (collaborative control, shared control) & exoskeleton; tactical validation & evidence presentation & state governance/control & purchasing behavior; dynamic pricing & custom configuration & research activities \\
            
            
            \textbf{Optimize \mbox{(Level~V)}} & training design to minimize stress/load & productivity & team composition & hiring processes & product & minimize muscular energy; minimize pre-operative preparation time; medication & predictive fee optimization; reduce healthcare costs; improve procurement margins & select/optimize learning methods/concepts & driving modes (e.g., goal: well-being) & performance; tactics & legal judgments & legislation & profit/revenue; supply chains; product offerings; shelf attractiveness; energy saving & evacuation planning & human-centered research developments \\
        \end{tabular}
        } 
        
        \vspace{1em}
        
        
        \vspace{1em}

        \tiny
        \begin{tabular}{r l}
            \textbf{Denoted Stakeholder} & \textbf{Further Applicable Stakeholders} \\
            \hline
            \textbf{Me} &  Patient, Driver, Worker, Musician, Person with Disability, Person with a Handicap, Non-auditory Person, Athlete, Hobby Sportsman, etc. \\
            \textbf{Work Planner} &  Work/Shift Planner \\
            \textbf{Shift Supervisor} & Work Preparation \\
            \textbf{Product Manufacturer} & Engineer, Medical Technology Manufacturer, Pharmaceutical Manufacturer, Product Designer, Architect, Fashion Producer/Designer \\
            \textbf{Trade} & Online Marketplace, Advertising Agency, Social Network, Public Institution, Business (Retail Store), Energy Supplier, Amusement Park, Owner of HDT Infrastructure \\
            \textbf{Management} & HR Department, Job Exchange, Job Center \\
            \textbf{Psychologist} & Physiotherapist, Caregiver, Doctor, Surgeon \\
            \textbf{Vehicle} & (semi-) autonomous vehicle \\
            \textbf{Military} & Relief Organizations \\
            \textbf{Politician} & Government, State, City \\
        \end{tabular}

    \end{minipage}
} 

\end{sidewaystable}

\clearpage

\end{document}